\documentclass{acm_proc_article-sp}

\usepackage{url}
\usepackage{amsmath}
\usepackage{verbatim}
\usepackage[ruled,linesnumbered,noend,noline,inoutnumbered]{algorithm2e}
\usepackage{graphicx}
\usepackage{paralist}

\newcommand{\Dfn}[1]{\textbf{\emph{#1}}}
\newcommand{\textcode}[1]{\texttt{#1}}
\newcommand{\ToDo}[1]{\emph{[\textbf{To Do:} #1]}}

\newtheorem{lesson}{Lesson}

\begin{document}
%



\title{Relationship-Based Access Control for OpenMRS}

%
%
%
%
%

\numberofauthors{2} 
%
\author{
%
%
\alignauthor
Syed Zain Raza Rizvi\quad Philip W. L. Fong\\
\affaddr{University of Calgary}\\
\email{\{szrrizvi, pwlfong\}@ucalgary.ca}
\alignauthor
Jason Crampton\qquad James Sellwood\\
\affaddr{Royal Holloway, University of London}\\
\email{jason.crampton@rhul.ac.uk}\\
\email{james.sellwood.2010@live.rhul.ac.uk}
}

\maketitle

\begin{abstract}
  Inspired by the access control models of social network systems,
  Relationship-Based Access Control (ReBAC) was recently proposed as a
  general-purpose access control paradigm for application domains in
  which authorization must take into account the relationship between
  the access requestor and the resource owner.  The healthcare domain
  is envisioned to be an archetypical application domain in which
  ReBAC is sorely needed: e.g., my patient record should be accessible
  only by my family doctor, but not by all doctors.

  In this work, we demonstrate for the first time that ReBAC can be
  incorporated into a production-scale medical records system,
  OpenMRS, with backward compatibility to the legacy RBAC mechanism.
  Specifically, we extend the access control mechanism of OpenMRS to
  enforce ReBAC policies.  Our extensions incorporate and extend
  advanced ReBAC features recently proposed by Crampton and Sellwood.
  In addition, we designed and implemented the first administrative
  model for ReBAC. In this paper, we describe our ReBAC
  implementation, discuss the system engineering lessons learnt
  as a result, and evaluate the experimental work we have undertaken.
  In particular, we compare the performance of the various
  authorization schemes we implemented, thereby demonstrating the
  feasibility of ReBAC.
\end{abstract}

\category{D.4.6}{Security and Protection}{Access Control}


\keywords{Medical records system, relationship-based access control,
authorization graph, authorization principal, administrative model}

\section{Introduction}

OpenMRS \cite{OpenMRS} is a production-scale, open-source electronic medical records
system that has been deployed in many countries, including South
Africa, Kenya, Rwanda, India, China, United States, Pakistan, the
Phillipines, etc.  Despite its tremendous success and wide deployment,
OpenMRS has a limitation in its access control mechanism, which is an
instantiation of \Dfn{Role-Based Access Control (RBAC)}. This
limitation is the topic of the following posting in the developer
forum \cite{OpenMRSPosting}.
\begin{quote}
  \emph{The RBAC system provides a reasonably robust mechanism for
    restricting access to system behaviours; however, we do not yet
    have a mechanism for restricting access to specific data (e.g.,
    you can see data for patient $X$, but not patient $Y$; or, you can
    see your patient's data except for specific lab results).}
\end{quote}
An interpretation of the above limitation is that, while it is
possible to restrict access of patient records to the role of doctors,
it is not possible to restrict access of \emph{my} patient record to
\emph{my} family doctor.  RBAC satisfies the access control
requirements of business domains in which data objects are ``owned''
by the organization, and thus all qualified personnel (i.e., of a
certain role) may be granted access.  In application domains in which
privacy is a concern, the data objects are sometimes ``owned'' by
individuals. There is now a need for finer-grained access control:
e.g., my patient record shall only be accessible by the clinicians who
are actually treating me.  That is, access is granted on the basis of
how the requestors are related to me.


The above access control challenge is one of the primary motivations
for the recently proposed \Dfn{Relationship-Based Access Control
  (ReBAC)} models.  Originally inspired by the access control models
of social network systems (e.g., Facebook), ReBAC grants access based
on how the access requester is related to the resource owner (e.g.,
friends, friends-of-friends).  This is in contrast with RBAC, in which
access is granted by considering the attributes of the requestor.
Fong \emph{et al.}~proposed a series of general-purpose ReBAC models
\cite{Fong:2011,Fong-Siahaan:2011,Bruns-etal:2012}, in which ReBAC is
envisioned to be applied to application domains other than social
computing, with the healthcare domain being an archetypical example.
While the idea of ReBAC has undergone a number of recent extensions in
the literature \cite{Cheng-et-al:2012a, Cheng-et-al:2012b,
  Aktoudianakis-et-al:2013, Crampton-Sellwood:2014, Fong-etal:2013,
  Location}, \emph{what remains to be seen is the adoption of ReBAC in a
production-scale system for an application domain other than social
computing.  And this is the gap we attempt to bridge by extending the
access control subsystem of OpenMRS to include an implementation of a
ReBAC model.}

In this paper, we report our experience of extending the access
control subsystem of OpenMRS with a ReBAC model.  The ``diff'' between
our extension and the original OpenMRS code base consists of 25,754
lines (with no context lines).  The extension involves 113 new files,
26 new database tables, and 15 web pages.  Our contributions are the
following.
\begin{compactenum}
\item We demonstrated for the first time that ReBAC can be
  incorporated into a production-scale medical records system, and
  did so with backward compatibility to the legacy RBAC mechanism.
\item We identified system engineering issues that one needs to
  address when one is to cleanly and efficiently implement ReBAC in a
  large system (\S \ref{sec-architecture}, \S\ref{sec-auth-graph},
  \S\ref{sec-request} and \S\ref{sec-eval}).
\item We adapted, extended and implemented the advanced features of
  ReBAC that were recently proposed by Crampton and Sellwood
  \cite{Crampton-Sellwood:2014}.  The implemented features include a
  generalization of social graphs called authorization graphs
  (\S\ref{sec-auth-graph}), a ReBAC analogue of roles called
  authorization principals (\S \ref{sec-auth-prin}), and a Unix-style
  authorization mechanism for authorization principals
  (\S\ref{sec-authorization-mechanism}).  Because OpenMRS supports a
  rich mechanism of privilege matching, the notions of authorization
  principals and authorization algorithms as proposed in
  \cite{Crampton-Sellwood:2014} must be either adapted (\S
  \ref{sec-auth-prin}) or extended (\S
  \ref{sec-authorization-mechanism}).  Our extensions involve the
  proposal of two semantics of authorization (strict-grant vs
  liberal-grant), as well as a highly efficient principal matching
  algorithm based on the idea of lazy evaluation (lazy-match).  What
  is pleasantly surprising is that, even after extensive adjustments
  in our implementation, the basic spirit of Crampton and Sellwood's
  design is preserved, thereby demonstrating the robustness of their
  proposals.\label{item-schemes}
\item We designed and implemented an administrative model for ReBAC
  (\S \ref{sec-admin}).  To the best of our knowledge, this is the
  first such implementation.
\item We empirically evaluated the performance of the
   various authorization schemes in item \ref{item-schemes} above
   (\S \ref{sec-eval}).
\end{compactenum}

\section{Related Work}

It has long been observed that the health domain requires an access
control model that takes into account the relationship between the
resource owner and the access requestor when an authorization decision
is made \cite{Beznosov,Rostad-Edsberg}.  That was partly the reason
that led to the proposal of an extension of OpenMRS to incorporate
parameterized roles \cite{Fischer-et-al:2009}.

Relationship-Based Access Control was a term coined independently by
Gates \cite{Gates:2007} and Carminati and Ferrari
\cite{Carminati-Ferrari:2009} to refer to a paradigm of access control
in which authorization decisions are based on whether the resource
owner and the access requestor are related in a certain way.
Initially, ReBAC was envisioned to be applied to the domain of social
computing.  A seminal work with this application in mind was that of
Carminati \emph{et al.} \cite{Carminati-etal:2009}.

Fong \emph{et al.} proposed a series of general-purpose ReBAC models
\cite{Fong:2011, Fong-Siahaan:2011, Bruns-etal:2012}, and advocated
the adoption of ReBAC for application domains outside of social
computing.  The health domain was envisioned to be an application
domain in which ReBAC is particularly suited.  ReBAC protection states
are social networks.  Modal logic and hybrid logic were proposed as
policy languages for specifying ReBAC policies \cite{Fong:2011,
  Fong-Siahaan:2011, Bruns-etal:2012}.

In UURAC (user-to-user relationship-based access control)
\cite{Cheng-et-al:2012a}, a policy is specified in a regular
expression-based policy language.  Access is granted if the resource
owner and the access requestor are connected by a path made up of a
sequence of edge labels satisfying the regular expression.  An
algorithm for finding a path that honors the regular expression is
formulated.  In a subsequent work \cite{Cheng-et-al:2012b}, the 
protection states were extended to track relationships between user
and resources (U2R) as well as between resources and resources (R2R).
Another innovation is the provision for multiple policies to be
applicable to the protection of a resource, and the design of conflict
resolution policies (conjunctive, disjunctive and precedence) to
arbitrate authorization decisions.  The work proposes to employ ReBAC
to regulate administrative activities, but does not provide details on
how that is achieved.  The administrative actions we proposed in \S
\ref{sec-admin} has clear semantics of how they are protected by
security preconditions, and how their executions affect the protection
state.

In \cite{Fong-etal:2013}, a temporal dimension is
introduced into ReBAC, so that access control policies require
entities to be related in a certain way in the past. The goal of this
extension is to support the expression of social contracts in online
communities. In \cite{Location}, ReBAC is extended to account for
geo-social network systems, and the hybrid logic policy language is
extended to impose relationship constraints over people located in a
certain geographical neighbourhood.

Crampton and Sellwood recently proposed a series of extensions to
ReBAC \cite{Crampton-Sellwood:2014}.  The protection state is an
authorization graph that tracks relationships among users, resources,
as well as other abstract entities relevant to access control (e.g.,
groups, roles, etc).  They also proposed a ReBAC analogue of roles
called authorization principals.  The run-time semantics of
authorization principals are specified through path conditions, a
language akin to regular expressions.  An XACML-style conflict
resolution mechanism was proposed to arbitrate authorizations when the
access requestor is associated with authorization principals that
grant conflicting authorizations.  A UNIX-inspired authorization
procedure serves as a framework for binding these technologies
together.  Our implementation has adopted the ideas of authorization
graphs, authorization principals, and the UNIX-style authorization
procedure.  Yet we employ hybrid logic rather than path conditions for
specifying the denotation of authorization principals. Detailed
comparison with \cite{Crampton-Sellwood:2014} will be given in the
rest of this paper.

It has long been recognized that any practical access control system
must provide ways to modify the authorization state or policy
\cite{HRU, sandhu1999arbac97}. While administrative access control
models, which control modifications to policies, have been widely
studied for the protection matrix and RBAC, this is a relatively
unexplored area in the context of ReBAC.  Fong's ReBAC model allows
for changes to the protection state through the use of contexts
\cite{Fong:2011}, but we are not aware of any implementation of
administrative features for ReBAC.

\section{R\MakeLowercase{e}BAC Goes Open Source}

This section reviews the background materials needed for understanding
the rest of the paper.

\subsection{An Overview of ReBAC}

In a series of papers \cite{Fong:2011, Fong-Siahaan:2011,
  Bruns-etal:2012}, Fong \emph{et al.}~proposed a general-purpose
access control model for Relationship-Based Access Control (ReBAC).
This work is mainly based on the variant of the model discussed in
\cite{Bruns-etal:2012}.

The protection state of ReBAC is an edge-labelled, directed graph:
directed edges represent interpersonal relationships, and each edge is
labelled with a relation identifier to signify the type of relation
(e.g., \textsf{patient-of}).  In the original conception of ReBAC,
vertices represent users, and thus the protection state is a social
network.  (In \S \ref{sec-auth-graph}, we follow the proposal of
\cite{Crampton-Sellwood:2014}, and generalize the social network to an
authorization graph.)

A \Dfn{graph predicate} determines whether particular conditions,
relating the vertices in a graph, hold or not.  We might, for example,
define a predicate that returns true if two vertices are connected by
an edge having a particular label.  More formally, a \Dfn{graph
  predicate of arity $k$} is a Boolean-valued function $\mathit{GP}(G,
x_1, \ldots, x_k)$, where $G$ is a graph that defines the current
protection state of the system, and each $x_i$ is a vertex in $G$.
$\mathit{GP}$ evaluates to $1$ if and only if $(x_1, \ldots, x_k)$
belongs to a $k$-ary relation defined over $G$. A graph predicate
of arity $2$ is said to be a \Dfn{relationship predicate}.
The relationship predicate $\textsf{friend-of-friend}(G, x_1, x_2)$,
for example, returns $1$ iff there exists a path of length two or less
in $G$ connecting $x_1$ and $x_2$, where both
edges in the path are labelled with the relationship type \textsf{friend}.


A ReBAC policy has the form ``\emph{grant access to $r$ if
  $\mathit{RP}$ evaluates to $1$}'', where $r$ is a resource and
$\mathit{RP}$ is a relationship predicate \cite{Bruns-etal:2012}.  An
access request has the form $(v, r)$, where user $v$ wishes to access
resource $r$.  Access is permitted if $\mathit{RP}(G, u, v)$ evaluates
to $1$, where $u$ is the owner of resource $r$.  (In \S
\ref{sec-auth-prin}, we adopt a recent idea due to
Crampton and Sellwood \cite{Crampton-Sellwood:2014},
and formulate ReBAC policies in terms of authorization principals
--- a ReBAC analogue of roles.)


A graph predicate $\mathit{GP}(G, x_1, \ldots, x_k)$ (with a
relationship predicate as a special case) can be syntactically
specified as a Hybrid Logic formula $\phi$ \cite{Bruns-etal:2012} with
$k$ free variables.\footnote{ More specifically, a graph predicate of
  arity $k$ can be represented by a hybrid logic formula $\phi$ with
  $k$ free variables, such that $\phi$ is a Boolean combination of
  \Dfn{anchored formulas}.  Each anchored formula is one in which the
  top-level operator is $@_x$, where $x$ is one of the free variables.
  This is a generalization of the syntactic restriction adopted in
  \cite{Bruns-etal:2012} for relationship predicates (i.e., arity 2).
} A local model checker is an algorithm that takes as input (i) a
hybrid logic formula $\phi$ with $k$ free variables, (ii) a protection
state $G$, and (iii) $k$ vertices $v_1$, \ldots, $v_k$, and then
decides whether the $k$-ary graph predicate represented by $\phi$ is
satisfied by $v_1$, \ldots, $v_k$ in $G$.

In the rest of this paper, knowledge of hybrid logic is not necessary
for appreciating the contributions of this work.  Nevertheless,
examples of hybrid logic formulas will be shown to convey the realism
of our design.  Readers who are unfamiliar with hybrid logic can
safely skip those examples.

\subsection{The ReBAC Java Library}

A reusable Java library of ReBAC technologies was released under
open-source terms \cite{ReBACLib}.  The library was developed and
maintained separately from OpenMRS.  The library was also packaged as a
Maven module for easy integration with large projects.

A main feature of the ReBAC library is the implementation of a local
model checker for the hybrid logic policy language of
\cite{Bruns-etal:2012}.  This model checker is a cornerstone of the
authorization mechanism in our OpenMRS extension, allowing us to
determine membership in authorization principal (\S
\ref{sec-auth-prin}), as well as to test if an administrative action
is enabled and/or applicable (\S \ref{sec-authorization-mechanism}).

Recall that the inputs to a local model checker include a graph and
the abstract syntax tree (AST) of a hybrid logic formula.  To allow
the model checker to interoperate with different representations of
graphs, we have defined an abstract interface for graphs.  For
example, in the case of OpenMRS, relationship edges may come from
three different sources (\S \ref{sec-auth-graph}).  A concrete class
that makes appropriate queries to check for existence of each of the
three kinds of relationships will implement the abstract graph
interface, thereby allowing the model checker to interoperate with OpenMRS.

Similarly, abstract interfaces are declared for the AST nodes of
hybrid logic formulas.  This allows the model checker to work with
different representations of hybrid logic formulas.  One may ask why
there is a need for different representations of hybrid logic
formulas.  A motivating example comes from OpenMRS.  In OpenMRS, all
data objects are stored as persistent objects (via Hibernate
\cite{Hibernate}).  That includes AST nodes of hybrid logic formulas.
Consequently, there are concrete representational demands on how AST
node classes are declared (e.g., must be a subclass of a certain
superclass).  Declaring abstract interfaces for AST nodes allows our
model checker to interoperate with such representational
idiosyncrasies.

Other features of the library include an XML parser for hybrid logic
formulas that are stored as XML files.

\section{Architecture of O\MakeLowercase{pen}MRS}
\label{sec-architecture}

This section introduces the architecture of OpenMRS, and explains
how ReBAC is built on top of this architecture.

\subsection{Interposition via AOP}

Our ReBAC implementation is based on the source code of OpenMRS
1.10.\footnote{The latest stable version of OpenMRS is 2.0, released
  on February 26, 2014.}  OpenMRS is built on the Spring Framework,
which is a Java-based web application framework \cite{Spring}.  Core
functionalities of OpenMRS are exposed as \Dfn{service-layer methods}
on the web application server.  The HTML pages invoke the
service-layer methods in order to query application data or alter
application state.  Access control is achieved by limiting access to
the service-layer methods.

Each service-layer method is annotated with either one of two kinds of
guard.\footnote{Annotation is achieved via the Java custom annotation
  mechanism \cite[\S 9.7]{JavaLangSpec}.}  Intuitively, a guard is a
specification of privilege requirements that must be satisfied by the
requestor in order for the invocation of the service-layer method to
be allowed.
\begin{compactenum}
\item \textcode{one-of($P$)}: Here $P$ is a set of privileges (i.e., 
 positive permissions). The intended meaning is that the invoker of
 this method must have been granted one of the privileges in $P$
 in order for method invocation to be authorized.
\item \textcode{all-of($P$)}: The invoker must have all of the
  privileges in $P$.
\end{compactenum}
Formally, we write $Q \models g$ for a set $Q$ of privileges and a
guard $g$ whenever $Q$ \Dfn{satisfies} $g$ in the following sense:
\begin{align*}
Q \models \textcode{one-of(}P\textcode{)} & 
  \quad\text{iff}\quad P\cap Q \neq \emptyset \\
Q \models \textcode{all-of(}P\textcode{)} &
 \quad\text{iff}\quad P \subseteq Q 
\end{align*}
Intuitively, if a requestor $u$ has been ``\emph{granted}'' a set $Q$
of privileges, and $Q \models g$, where $g$ is the guard of the method
that $u$ attempts to invoke, then invocation is authorized.  (As we
shall see in \S \ref{sec-authorization-mechanism}, there are two ways
to interpret the word ``\emph{granted}'', thereby yielding two
authorization semantics.)

Spring uses aspect-oriented programming (AOP)
\cite{Kiczales-etal:1997} to implement interposition of authorization
checks.  The original authorization checking code is implemented as an
``advice'' (more precisely, a ``before advice'') that is ``weaved''
into the entry point of each service-layer method, thereby introducing
additional behaviour on method entry.  Thus every method invocation is
intercepted by the RBAC authorization mechanism.  To implement ReBAC,
we introduced an additional authorization advice.  The ReBAC
authorization advice is an ``around advice'', which introduces
additional behaviour at both the entry and exit of a method.
Consequently every method invocation as well as method return is
intercepted by the ReBAC authorization mechanism.

\begin{lesson}
  Physically localizing all authorization checks in an identifiable
  code unit (e.g., module, reference monitor, aspect, etc) greatly
  eases the extension of the authorization mechanism to incorporate
  ReBAC.
\end{lesson}

In fact, the above lesson applies generally to all software systems
that anticipate future evolution in their authorization mechanisms
(incorporating ReBAC is but one possible evolution), and we have very
positive experience with AOP in this regard.

\subsection{Combining RBAC and ReBAC}

Unmodified, OpenMRS enforces a Role-Based Access Control (RBAC) model
\cite{Sandhu-etal:1996}, although the notion of sessions is not
implemented.  That is, all roles assigned to a user are activated when
the user logs into the system.  The likely reason is that the notion
of role activation is probably too exotic for medical professionals,
and the extra step of role activation in every log-in attempt would
degrade care delivery efficiency. It has also been pointed out that
the support for sessions is not essential to core RBAC implementations
in certain application domains \cite{CritiqueANSI}.

As discussed above, the original RBAC authorization checks are
implemented as an advice. We implemented ReBAC authorization checks as
a separate advice.  The configuration is that the RBAC authorization
checks are conducted first, and only when access is granted by RBAC
will the ReBAC authorization checks be conducted. In summary, access
is granted when both the RBAC and ReBAC mechanisms authorize access.
We have also tailored configuration files in such a way that system
administrators who do not use the new ReBAC features will not observe
any difference between the original implementation and the extended
one.
\begin{lesson}[Backward Compatibility]
Care must be taken to ensure that ReBAC features are backward
compatible with the legacy access control model of the system.
\end{lesson}

Crampton and Sellwood proposed a way of ``encoding'' RBAC in their
extended ReBAC model \cite{Crampton-Sellwood:2014}.  This suggests an
alternative means for integrating ReBAC and RBAC: implement only a
ReBAC model, and simulate RBAC with ReBAC.  Such an approach would be
particularly fitting if the software application is written from
scratch with a requirement to support both access control models.

\subsection{Protection and Application State}
\label{sec-prot-state}

In an application with a traditional access control model (e.g.,
RBAC), the protection state (e.g., role hierarchy, user-role
assignment, etc) of the system is separate from its application state
(i.e., application data).  This is true of the original architecture
of OpenMRS.

In social computing systems, however, the above is not necessarily
true.  For example, the interpersonal relationships articulated by
users in a social network system is both application data and part of
the protection state: authorization is granted based on the
relationship between the resource owner and requestor.  Inspired by
social computing applications, ReBAC inherits this overlapping of
protection and application state.

The above overlap is also present in the ReBAC extension of OpenMRS.
Included in a patient record is a set of users (e.g., family members)
related to the patient, as well as their relationships.  This, for
example, allows clinicians to anticipate hereditary conditions, or to
identify compatible blood, organ and tissue donors.  These
relationships obviously belong to the application state of OpenMRS.
Yet, as we shall see below, ReBAC authorization checks also make use
of such relationships when an authorization decision is computed.
That is, these relationships constitute part of the protection state.

The above overlap creates something of a dilemma.  In the original
OpenMRS architecture, patient relationships are accessible only via
service-layer methods, thereby ensuring complete mediation.  Yet, the
ReBAC authorization advice also needs to access patient relationships.
The advice will therefore need to invoke service-layer methods in
order to access the relationships.  As invocations of service-layer
methods are intercepted by the authorization advice, this
inevitably leads to an infinite loop.

All patient data, including patient relationships, are stored as
Hibernate persistent objects \cite{Hibernate}.  These persistent
objects are made accessible via Data Access Objects (DAOs).  To break
the infinite loop, we created direct access paths to patient
relationships by configuring DAOs specifically for the ReBAC
authorization advice, so that the latter may access patient
relationships without mediation of authorization checks.  The above
experience leads to the articulation of the following general lesson
for ReBAC systems.
\begin{lesson}[Application and Protection State] \ \linebreak Data
  belonging to both the application and protection state of a system
  must be held in a data store which exposes two Application
  Programming Interfaces (APIs).  One is mediated by authorization
  checks, the other is not.  The mediated API is invoked by users,
  while the unmediated one is utilized internally for authorization.
\end{lesson}

\section{Authorization Graph}
\label{sec-auth-graph}

In the early conception of ReBAC \cite{Bruns-etal:2012}, the
protection state is a social network of users: an edge-labelled,
directed graph in which vertices represent users and edges model their
interpersonal relationships.  Crampton and Sellwood proposed an
extension of ReBAC in which the protection state is an
\Dfn{authorization graph} \cite{Crampton-Sellwood:2014}. The vertices
model not only users, but also resources as well as other entities
that are relevant to access control (e.g., groups).  The edges capture
relationships among users, objects and the aforementioned entities.
Our ReBAC adaptation of OpenMRS implements the idea of authorization
graphs.

When one applies ReBAC to an enterprise application domain (i.e., a
domain other than social computing), a frequently raised question is:
Where do the relationships come from?  This rest of this section
reports our answer to this question, as shaped by our experience with
OpenMRS.

In OpenMRS, domain objects are all instances
of the root class \textcode{BaseOpenmrsObject},
which has two subclasses \linebreak \textcode{BaseOpenmrsData}
and \textcode{BaseOpenmrsMetadata}.  The instances
of \textcode{BaseOpenmrsData} include users, patient 
records and their components, etc.  Therefore, we
take all instances of \linebreak \textcode{BaseOpenmrsData}
as the vertices of the authorization graph.

The authorization graph tracks binary relationships among instances of
\textcode{BaseOpenmrsData}.  During our development of the ReBAC
extensions for OpenMRS, we identified three categories of
relationships.
\begin{compactenum}
\item \Dfn{User-managed relationships}.  These are relationships that
  are explicitly articulated and managed by end users.  An example is
  friendship in Facebook.

  As we mentioned in \S \ref{sec-prot-state}, OpenMRS enables a
  clinician to document in a patient record the relatives of the
  patient.  These interpersonal relationships are considered part of
  the authorization graph.  More specifically,
  \textcode{BaseOpenmrsData} has a subclass \textcode{Person}.
  Recorded interpersonal relationships between instances of
  \textcode{Person} are considered to be edges in the authorization
  graph.

\item \Dfn{System-induced relationships}.
  The data structures of the system may contain
  relationships that are relevant to authorization.
  Examples include organizational structures, object 
  ownership, object containment and provenance relationships.
  End users are not allowed to directly manipulate these
  relationships.

  In our ReBAC adaptation of OpenMRS, we have created an extension
  mechanism for administrators to introduce new system-induced
  relationships.  Specifically, a system-induced binary relation is
  implemented as a Java class that performs queries into the run-time
  data structures of OpenMRS.  Such a class implements the
  \textcode{ImplicitRelationIdentifier} interface, which defines a
  standard calling convention for performing relationship queries.  At
  run-time, such a class will be dynamically loaded into the Java
  Virtual Machine,
   an instance of that class is
  created, and an appropriate method of that instance will be invoked
  when the authorization mechanism needs to check the system-induced
  relation.  The administrator can install an extension class for each
  type of system-induced relationship.

  In our ReBAC adaptation of OpenMRS, a system-induced relation
  relates instances of \textcode{BaseOpenmrsData}, meaning that such
  relationships are not only among users, but they may also relate
  resources to resources, or persons to resources.  As an example of
  the last case, we implemented resource ownership (\textsf{owner}) as
  a system-induced relation, relating a resource to its owner(s).
  
\item \Dfn{Access control relationships}.  There are relationships
  that belong solely to the protection state: they are tracked solely
  for the purpose of access control, and have no relevance to the
  business logic of the application.  Examples of access control
  relationships include role or group membership, records of
  access events (e.g., for implementing history-based policies,
  as in \cite{Fong-etal:2013}), etc.

  In our ReBAC adaption of OpenMRS, access control relationships are
  defined among instances of the \textcode{Person} class.  Manually
  adding or removing access-control edges in the authorization graph
  is an error-prone step.  To reduce the cognitive burden of users, we
  have implemented an administrative model for ReBAC, thereby
  supporting a principled way for adding or removing access control
  relationships.  See \S \ref{sec-admin} for details.

  For example, say the family doctor of a patient may refer the
  patient to a specialist.  Such a capability is only allowed if
  patient and a clinician are related by an access control relationship
  \textsf{family-doctor}.  Once the referral is confirmed, the patient
  and the specialist will be related by the access control
  relationship \textsf{referred-clinician}, thereby enabling the
  specialist to access the patient's record.
\end{compactenum}
\begin{lesson}
  In a ReBAC system, relationships come from three sources.  Some
  relationships belong purely to the protection state (i.e., access
  control relationships): these are managed by system administrators.
  Other relationships are shared between the application state and the
  protection state.  This latter kind may be further classified into
  (i) relationships that are explicitly articulated and managed by end
  users, and (ii) relationships that are induced by the system data
  structure (and thus cannot be manipulated directly by users and
  administrators).
\end{lesson}

\section{Access Requests}
\label{sec-request}

The ReBAC authorization advice needs three pieces of information to
compute an authorization decision: (a) the resource $r$ to which
access is required, (b) the user $u$ who wishes to have access (aka
the ``requestor''), and (c) the guard $g$ of the service-layer method
being invoked.  Therefore, an access request in OpenMRS is
characterized by a triple $(r, u, g)$.

The ReBAC authorization advice can discover the identity of the
requestor ($u$) and the service-layer method that is being
invoked.\footnote{The requestor can be identified by calling a public
  static method of the \textcode{Context} class in OpenMRS.  The ReBAC
  authorization method is passed an argument of type
  \textcode{MethodInvocation}, which in turn provides access to the
  identity of the service-layer method that is being invoked.}  Using
the Java Reflection API, the ReBAC authorization advice can then
extract the guard ($g$) of the service-layer method. The last
component of the access request, namely the resource $r$, is not
directly available. OpenMRS was originally designed to use RBAC for
authorization, and that explains why the identity of the resource is
not explicitly made available for the authorization mechanism.  In the
following, we discuss how the requested resource $r$ is identified in
a systematic manner for the ReBAC authorization advice.

The ReBAC authorization advice has access to the arguments that are
passed to the service-layer method, as well as the return value of
that invocation.  Depending on the kind of service-layer method, the
target resource may be either (a) an argument or (b) the return value.
There are two kinds of service layer methods in OpenMRS:
\begin{compactenum}
\item A \Dfn{setter} method is one that operates on a given resource,
  which appears as one of the method arguments.  That is, a setter
  produces side effects on the application state.  The argument for
  which side effect is targeted is the resource that requires access
  control.
\item A \Dfn{getter} method retrieves patient information (e.g.,
  searching for the records of all patients with a given family name).
  The return value of a getter method is either (a) a single piece of
  patient information, or (b) a collection or a map of patient
  information.  In the former case, the returned patient datum is the
  resource that requires access control, and in the latter case, every
  returned patient datum requires access control.
\end{compactenum}
In the original design of OpenMRS, a naming convention is adopted to
differentiate getter and setter methods, but there is no way for the
ReBAC authorization advice to recognize which argument of a setter
requires access control.


To address the above problem, we designed a custom annotation
\textcode{@Resource} for identifying (a) whether 
a service-layer method is a setter or a getter, and (b)
the target resource for each kind.  In the case of setter methods,
the \textcode{@Resource} annotation can be applied to a method
parameter to indicate that that parameter corresponds to a protected
resource.  \textcode{
\begin{tabbing}
$T$ $m$($T_1$ $x_1$, @Resource $T_2$ $x_2$, $T_3$ $x_3$) \{ \ldots\ \}
\end{tabbing}
}
\noindent The \textcode{@Resource} annotation is applied above to
explicitly declare that the parameter $x_2$ of method $m$ is a
controlled resource.  We systematically annotated the setter methods
in the OpenMRS code base using the above annotation.

When the authorization advice is invoked, it employs the Java
Reflection API to discover if any of the parameters of the invoked
method is annotated by \textcode{@Resource}.  If so, then it will pass
the request $(r, u, g)$ through the authorization procedure, where $r$
is the value of the annotated parameter.  Invocation of the method is
only granted if authorization is successful.

Similarly, the \textcode{@Resource} annotation can also be applied to
the method as a whole to declare that the method is a getter and thus
the return value requires access control.\footnote{The annotation of
getter methods is
  not absolutely necessary, as the above-mentioned naming convention
  already 
identifies getter methods. The annotation is performed as a
convenience for the ReBAC authorization advice.  The annotation of
setter methods, however, is necessary in order for the
ReBAC authorization advice to function properly. }  \textcode{%
\begin{tabbing}%
@Resource $T$ $m$($T_1$ $x_1$, $T_2$ $x_2$, \ldots) \{ \ldots\ \}%
\end{tabbing}%
}%
\noindent Again, we systematically annotated the getter methods in the OpenMRS
code base using the above annotation.

Before an invoked method returns, the ReBAC authorization advice will
check if the method has the \textcode{@Resource} annotation.  If so,
it will perform authorization checks on the return value.  If the
return value is a single piece of patient information $r$, then the
request $(r, u, g)$ will be subject to the authorization procedure,
and a security exception will be raised if authorization fails.
Otherwise, the return value is either a collection or a map of patient
information.  For every member $r$ in the returned collection
(resp.~map), the request $(r, u, g)$ will be subject to authorization
check.  A collection (resp.~map) containing only those $r$s that pass
authorization will be returned.

\begin{lesson}
  The legacy authorization subsystem of some applications may not have
  direct access to both the requestor and the resource of an access
  request.  A ReBAC extension of such applications will need to
  provide means for run-time identification of these two entities.
\end{lesson}

\section{Authorization Principals}
\label{sec-auth-prin}

In an early conception of ReBAC \cite{Bruns-etal:2012}, a ReBAC policy
has the form ``\emph{grant access to $r$ if $\mathit{RP}$}'', where
$r$ is a resource and $\mathit{RP}$ is a relationship predicate.
There are two limitations to this design. First, access to resource
$r$ may be performed via many different forms of operations, and thus
finer grained access control based on permissions (i.e., privileges in
OpenMRS) is desirable. Second, there is no provision of permission
abstraction (i.e., such as roles in RBAC) to ease administration.  To
overcome these limitations, Crampton and Sellwood
\cite{Crampton-Sellwood:2014} proposed an extension of ReBAC that is
based on permission granting, and invented the notion of
\Dfn{authorization principals}, which could be seen as a ReBAC
analogue of roles, to ease administration.  In our ReBAC extension of
OpenMRS, we have adopted a variant of Crampton and Sellwood's
proposal.  In the following, we will first describe the scheme that we
actually implemented, and then discuss how it differs from the
original proposal of Crampton and Sellwood.

An authorization principal is defined via a principal matching rule of
the form $(\mathit{AP}, \mathit{RP})$, where $\mathit{AP}$ is the
identifier of the authorization principal, and $\mathit{RP}$ is a
relationship predicate.  Unlike a role in RBAC, in which membership in
a role is defined statically (via the user-role assignment relation
$\mathit{UA}$), the semantics of an authorization principal is
dynamic.  When a request to access resource $r$ is issued (at run
time), $\mathit{AP}$ denotes the set of users $u$ for which $r$ and
$u$ satisfy $\mathit{RP}$, the relationship predicate that is
associated with $\mathit{AP}$.  This notion of authorization
principals is actually familiar to us.  For example, in Unix, there
are three built-in authorization principals: ``owner'', ``group'',
``other'' (aka ``world''); in Facebook, there are four built-in
authorization principals: ``me'', ``friend'', ``friend-of-friend'',
``everyone''.

Note again that, in the original conception of ReBAC
\cite{Bruns-etal:2012}, the relationship predicate in a ReBAC policy
specifies a desired relation between the resource owner and the access
requestor.  In contrast, the relationship predicate in a principal
matching rule specifies a desired relation between the resource itself
and the requestor.  For example, the following principal matching
rule specifies the principal $\mathsf{treating}\text{-}\mathsf{clinician}$.
\begin{multline*}
   \Big( \mathsf{treating}\text{-}\mathsf{clinician},\\
     @_{\mathsf{resource}} \langle
        \mathsf{owner}
     \rangle \big(
       \langle 
          \mathsf{family}\text{-}\mathsf{doctor}
       \rangle \mathsf{requestor}
       \lor \mbox{}\\
       \langle 
          \mathsf{referred}\text{-}\mathsf{clinician}
       \rangle \mathsf{requestor}
     \big)
   \Big)
\end{multline*}
The rule says that the requestor is a treating clinician if she is
either the family doctor or a referred specialist of the resource's
owner.  Note that the two free variables \textsf{resource} and
\textsf{requestor} identifies the two parameters of the relationship
predicates.

In our implementation, there is only one principal matching rule for
each authorization principal $\mathit{AP}$: i.e., the principal
matching rule defines a functional mapping from authorization
principals to their corresponding relationship predicates.  We write
$\mathit{RP}_{\mathit{AP}}$ for the relationship predicate of the
authorization principal $\mathit{AP}$.

Permission abstraction is achieved by authorization rules of the form
$(\mathit{AP}, P)$, where $\mathit{AP}$ is the identifier of an
authorization principal, and $P$ is a set of privileges.  The meaning
is analogous to the permission assignment relation $\mathit{PA}$ in
RBAC.  That is, at run time, the members of authorization principal
$\mathit{AP}$ is granted permissions in $P$.

In our implementation, there is only one authorization rule
for each authorization principal.  We write $P_{\mathit{AP}}$
for the set of privileges granted to authorization principal
$\mathit{AP}$.

The scheme we implemented differs from the original proposal of
Crampton and Sellwood in the following manners.
\begin{compactitem}
\item Crampton and Sellwood use a formalism called path conditions to
  specify the relationship predicate $\mathit{RP}$.  In our
  implementation, $\mathit{RP}$ is specified via a hybrid logic
  formula.  Path conditions and hybrid logic have incomparable
  expressiveness. There are certain relationship predicates that are
  expressible in hybrid logic but not path conditions, and vice versa.
  Extending our implementation to accommodate other specification
  formalisms for relationship predicates is a modular task.
\item In Crampton and Sellwood's proposal, an authorization rule may
  grant either positive or negative permissions (i.e., allow or deny).
  Complying to the original design of OpenMRS, our implementation
  supports only positive permissions.  Without negative permissions,
  the conflict resolution strategies proposed in
  \cite{Crampton-Sellwood:2014} are not needed and thus not
  implemented.  Extension of our implementation to accomodate negative
  permissions and conflict resolution is a tractable endeavour.
\item An authorization rule of Crampton and Sellwood has an explicitly
  specified scope of applicability.  Specifically, a rule is either
  applicable to all resources, or it is applicable only to a specific
  resource $r$.  Our implementation supports only the first 
  possibility (applicable to all resources).
\end{compactitem}

A number of user interface elements have been introduced to ease the
administration of authorization principals and privilege assignment.
First, the specification of principal matching rules and the
specification of authorization rules are performed in two separate web
pages.  Each web page is protected by separate privileges.  This
separation of duty allows a different group of administrators to be
responsible for specifying each kind of rules.  Second, we have
developed a Javascript-based structure editor for specifying Hybrid
Logic formulas (e.g., as relationship predicates in principal 
matching rules).

\section{Authorization Mechanism}
\label{sec-authorization-mechanism}

Inspired by the UNIX access control model \cite{Crampton:2008},
Crampton and Sellwood proposed an authorization mechanism for
determining when a request is to be granted. We adapted their proposal
for OpenMRS.  Given an access request $(r, u, g)$ directed against a
protection state (i.e., an authorization graph) $G$, an authorization
principal $\mathit{AP}$ is said to be \Dfn{enabled} iff
$\mathit{RP}_{\mathit{AP}}(G, r, u) = 1$.  Intuitively, the requestor
$u$ is a member of the enabled principals for the present access
request.  Thus, requestor $u$ is granted the privileges in
$\mathit{P}_{\mathit{AP}}$, for each enabled principal $\mathit{AP}$.
Such privileges are then used for satisfying the privilege requirement
of guard $g$.

The presence of guards of the form \textcode{all-of($P$)} present
ambiguities in the precise manner in which authorization should be
conducted.  We therefore extend the proposal of Crampton and Sellwood
by differentiating between two semantics of authorization.
\begin{compactenum}
\item \Dfn{Liberal-grant semantics}.  
\begin{compactitem}
\item Let $\mathcal{E}$ be the set of all enabled principals.
\item Let $Q = \bigcup_{\mathit{AP} \in \mathcal{E}} P_{\mathit{AP}}$.
  That is, $Q$ is the set of all privileges that are granted by at
  least one enabled principal.
\item Authorization is granted iff $Q \models g$.
\end{compactitem}
In liberal-grant authorization, the privileges required by $g$
may come from any enabled principals.  The assumption is that
the requestor $u$ can simultaneously ``be'' all the enabled
principals.
\item \Dfn{Strict-grant semantics}.
\begin{compactitem}
\item Let $\mathcal{E}$ be the set of all enabled principals.
\item Authorization is granted iff there exists $\mathit{AP} \in
  \mathcal{E}$ such that $P_\mathit{AP} \models g$.
\end{compactitem}
In strict-grant authorization, the privileges required by $g$ must
originate from only one enabled principal.  The idea is that the
privilege requirements of $g$ are satisfied only if there is an
enabled principal who can ``single-handedly'' satisfy it.
\end{compactenum}
The two semantics produce identical behaviour if the guard $g$ is of
the form \textcode{one-of($P$)}.\footnote{If $g =
  \textcode{one-of(}P\textcode{)}$, then $P \cap (\bigcup_{\mathit{AP}
    \in \mathcal{E}} P_\mathit{AP} ) \neq \emptyset$ iff there exists
  $\mathit{AP} \in \mathcal{E}$ such that $P \cap P_\mathit{AP} \neq
  \emptyset$.  That is, the two semantics agree in their authorization
  decisions.} They differ in behaviour only if the guard is of the
form \textcode{all-of($P$)}.\footnote{ Suppose $g =
  \textcode{all-of(}\{p_1, p_2\}\textcode{)}$.  Suppose further
  $\mathcal{E} = \{ \mathit{AP}_1, \mathit{AP}_2 \}$.  Say
  $P_{\mathit{AP}_1} = \{ p_1 \}$ and $P_{\mathit{AP}_2} = \{ p_2
  \}$. Then liberal grant semantics will allow access but strict grant
  semantics will deny access.  }  If a request is authorized in the
strict-grant semantics then it is authorized in the liberal-grant
semantics.\footnote{ Suppose strict-grant allows access. There exists
  $\mathit{AP} \in \mathcal{E}$ such that $P_\mathit{AP} \models g$.
  In that case, $\bigcup_{\mathit{AP} \in \mathcal{E}} P_\mathit{AP}
  \models g$ as well, since $\models$ is monotonic.  Thus, liberal
  grant allows access also.}

For each of the above semantics, we also developed two
implementations.
\begin{compactenum}
\item \Dfn{Eager-match strategy}.  This is the straightforward
  implementation of the two semantics, in which the set of all enabled
  principals is computed before an authorization decision is produced.
\item \Dfn{Lazy-match strategy}.  This is an optimized implementation
  of the two semantics.  The core idea is that the testing of
  relationship predicates during principal matching (i.e., determining
  which principals are enabled) is an expensive operation, and thus
  such checks should be avoided whenever possible.  This idea is
  materialized in two ways.  First, two principals may share the same
  relationship predicate.  There is no point re-evaluating the
  predicate for both principals. When we determine what principals are
  enabled, the same relationship predicate is evaluated only once.
  Second, rather than computing the set of all enabled authorization
  principals, they are computed one at a time, and only for the
  principals that are relevant.  If the privileges associated with a
  principal do not contribute to the satisfaction of the guard in
  question, it is ignored, and its relationship predicate is not even
  evaluated.  Otherwise, the principal is ``relevant'', and its
  relationship predicate is checked to see if the principal is
  enabled.  Whenever a relevant principal is found to be enabled, the
  authorization engine checks to see if the required privileges are
  already present.  If so, the search for enabled principals will be
  terminated.  In summary, this ``lazy'' evaluation strategy
  opportunistically eschews unnecessary computation.  The pseudocode
  listings for liberal-grant and strict-grant authorization using the
  lazy-match strategy are shown in Algorithms \ref{algo-liberal} and
  \ref{algo-strict} respectively.
\end{compactenum}
The two strategies produce the same authorization decision for any
given access request.

\begin{algorithm}[t]
let $P$ be such that $g$ is either $\textcode{all-of}(P)$
  or $\textcode{one-of}(P)$\;
$Q := \emptyset$\;
\ForEach{$\mathit{AP}$}{
\If{$(P_{\mathit{AP}} \setminus Q) \cap P \neq \emptyset$}{
\If{$\mathit{RP}_{\mathit{AP}}(G, r, u)$ has been evaluated}{
  reuse previous value\;
}
\Else{
  compute value\;
}
\If{value is true}{
  $Q := Q \cup P_{\mathit{AP}}$\;
  \If{$Q \models g$}{
    \Return{``allow''\,}\;
  }
}
}
}
\Return{``deny''\,}\;
\caption{Lazy-match, liberal-grant authorization of access request
  $(r, u, g)$ against authorization graph $G$.\label{algo-liberal}}
\end{algorithm}

\begin{algorithm}[t]
\ForEach{$\mathit{AP}$}{
\If{$P_{\mathit{AP}} \models g$}{
\If{$\mathit{RP}_{\mathit{AP}}(G, r, u)$ has been evaluated}{
  reuse previous value (which must be false)\;
}
\Else{
  compute value\;
}
\If{value is true}{
\Return{``allow''\,}\;
}
}
}
\Return{``deny''\,}\;
\caption{Lazy-match, strict-grant authorization of access
request $(r, u, g)$ against authorization graph $G$\label{algo-strict}.}
\end{algorithm}

We implemented a web interface for administrators to select between
liberal- or strict-grant semantics, and between eager- or
lazy-match strategy (i.e., four combinations).

\section{Administrative Actions}
\label{sec-admin}

Access control relationships in the authorization graph belong solely
to the protection state.  They are not application data. Their
existence serve only the purpose of protection.  One way of managing
such relationships will be to place the burden entirely on the system
administrators.  (In our implementation, we have administrative web
pages for administrators to manually add or delete edges of the
authorization graph.)  This, however, is not scaleable.  Imagine the
task of adding an edge in the authorization graph to indicate that the
family doctor of a patient is referring the patient to a cardiologist
(and thus the said cardiologist enjoys certain access rights that
other cardiologists do not have over the patient's records).  Such an
action is common in the daily operation of a health service. It is
completely impractical to go through the bottleneck of the system
administrators every time such a referral is made.  One way of making
this scaleable is to delegate this operation to qualified users (e.g.,
the family doctor in the example), so that the latter may add this
edge into the authorization graph.  Yet, manual addition and deletion
of edges can be error prone.  First, business logic may dictate that
multiple updates to the authorization graph must occur together (e.g.,
a person may have only one supervisor, and thus the addition of a new
supervisor edge must be accompanied by the deletion of an out-of-date
supervisor edge). If the user performs one update but forgets another,
then the integrity of the authorization graph cannot be
maintained. Second, business logic may dictate that an update can only
occur if the user performing the update is qualified to do so (e.g.,
referral can only be made by a family doctor).  Undisciplined updates
of the authorization graph overlooks such security requirements.

The primary design objective of administrative actions is to provide a
structured means for adding and removing access control relationships,
so that such tasks can be performed safely by users other than system
administrators.  In our design, the declaration of an administrative
action consists of the following components:
\begin{compactitem}
\item \textbf{Action identifier}: A unique name is used for
  identifying the administrative action.  For example, the
  referral action is identified by the identifier 
  ``\textsf{Referral}''.
\item \textbf{Enabling precondition}: Every administrative action is
  presumed to be performed by a user against a patient (e.g., a family
  doctor performing a referral for a patient).  So every
  administrative action has two \Dfn{primary participants}, namely,
  the user who performs that action and the patient to which the
  action is targeted.  The identifiers \textsf{user} and
  \textsf{patient} are used in the declaration for referring to
  the primary participants.

  Whether the action is \Dfn{enabled} (see below) depends on whether
  the user and the participants satisfy a certain relationship
  predicate.  Such a relationship predicate, called the enabling
  precondition, is specified as a hybrid logic formula with free
  variables \textsf{user} and \textsf{patient}.

  For example, the following hybrid logic formula can be
  used for requiring that referral can only be conducted
  by the family doctor of a patient.\footnote{This is only
  an illustration. We are fully aware that in real life
  it is not just the family doctor who can perform referral.}
\[
   @_{\mathsf{user}} \langle \mathsf{family}\text{-}\mathsf{doctor}
    \rangle \mathsf{patient}
\]
\item \textbf{Participants}: Other than \textsf{user} and
  \textsf{patient}, there may be other participants involved in the
  action.  They are called \Dfn{auxiliary participants}.  The
  participant list enumerates the identifiers to be used for referring
  to auxiliary participants in the rest of the declaration.

  In the example of referral, there is only one auxiliary 
  participant, ``\textsf{specialist}'', who is the specialist
  to which the referral is directed.
\item \textbf{Applicability precondition}: Whether the administrative
  action is considered \Dfn{applicable} (see below) depends on whether
  a certain condition holds among all the participants (both primary
  and auxiliary).  Such a condition is specified as a hybrid logic
  formula containing free variables that are \textsf{user},
  \textsf{patient}, as well as the identifiers listed in the
  participant list above.  The hybrid logic formula specifies
  a graph predicate of arity $\ell + 2$, where $\ell$ is the
  number of auxiliary participants.

  In the running example, we require that (a) the \textsf{specialist}
  is approved by the insurance company of the \textsf{user}, and (b)
  the \textsf{user} and the \textsf{specialist} must belong to the
  same health region.  The above conditions are captured by the
  following hybrid logic formula.
  \begin{multline*}
    \big(
    @_{\mathsf{patient}} \langle
    \mathsf{insurance} \rangle
    \langle \mathsf{approves}\rangle
    \mathsf{specialist} 
    \big) \land \mbox{}\\
    \big(
    @_{\mathsf{user}} \langle
    \mathsf{region} \rangle
   \langle -
    \mathsf{region} \rangle
    \mathsf{specialist}
    \big)
  \end{multline*}
\item \textbf{Effects}: The effects of an administrative action is a
  list of \Dfn{updates}.  Each update is of the form
  ``\textbf{add} $i(x,\,y)$'' or ``\textbf{del} $i(x,\,y)$''.
  Here, $i$ is a relation identifier (e.g., \textsf{supervisor-of}),
  and $x$ and $y$ are identifiers of participants (either primary or
  auxiliary).  The keywords \textbf{add} and \textbf{del} indicates
  whether the update is an edge addition or deletion.

  For example, the referral action has one update:
  \[
     \text{\textbf{add} $\mathsf{referred}\text{-}\mathsf{clinician}(\mathsf{patient},\,
     \mathsf{specialist})$}
  \]
\end{compactitem}
The enabling and applicability preconditions together specify the
security constraints that must be met in order for the \textsf{user}
to be allowed to perform the administrative action against the
\textsf{patient}.  The effects may involve updating multiple edges in
the authorization graph.  Grouping them together in one administrative
action ensures that the updates are either performed together, or not
at all.  This in turn ensures the integrity of the authorization
graph, and prevents errors on the part of the \textsf{user} who
performs the updates.

At run time, the following sequence of events occur,
which gives semantics to administrative actions.
\begin{compactenum}
\item When a user retrieves the record of a patient, the two primary
  participants are tested against the enabling precondition of every
  declared administrative action.  An action for which the enabling
  precondition is satisfied is said to be \Dfn{enabled}.
  The set of enabled actions is computed.
\item When the patient record is displayed, a tab showing
  the list of all enabled actions is made available to the user.
  The user may choose to perform any of the enabled actions.
\item When the user signals to perform an enabled action, the list of
  auxiliary participants (if any) will be displayed to the user.  The
  user must now instantiate each of the participants by selecting a person.
  This is facilitated by intelligent search features offered by
  OpenMRS.
\item \label{step-check} Once the participants are selected, both the
  enabling and applicability preconditions are checked.  The action is
  deemed \Dfn{applicable} if the check succeeds.  (We will explain
  below why the enabling precondition is checked again.)
\item If the action is applicable, then the effects of the action will
  be executed.  Note that deleting a non-existent edge is an error.
  Similarly, adding an edge that already exists is also considered an
  error.  Either all the updates are executed, or execution fails
  without any change to the authorization graph.
\end{compactenum}
Note that the execution of effects is an atomic operation: either all
updates are successfully executed, or else no update is performed.
This is achieved by the transaction manager.  Actually, the
transaction begins at step \ref{step-check} above.  Including the
check of both enabling and application preconditions into the
transaction prevents time-of-check-to-time-of-use (TOCTTOU) race
conditions \cite[\S 6.2.1]{Anderson}.  In addition, to prevent
unintended roll-back, the preconditions should be crafted in such a
way that the presence of an edge is verified in the preconditions if
it is deleted in the effects, and the absence of an edge is confirmed
in the preconditions if it is added in the effects.

To support policy engineering, we also developed administrative web
pages for users to build a library of reusable hybrid logic formulas.
Such formulas can be referenced in the declarations of administrative
actions.

\section{Performance Evaluation}
\label{sec-eval}

\begin{figure}
\centering
\includegraphics[width=.7\linewidth]{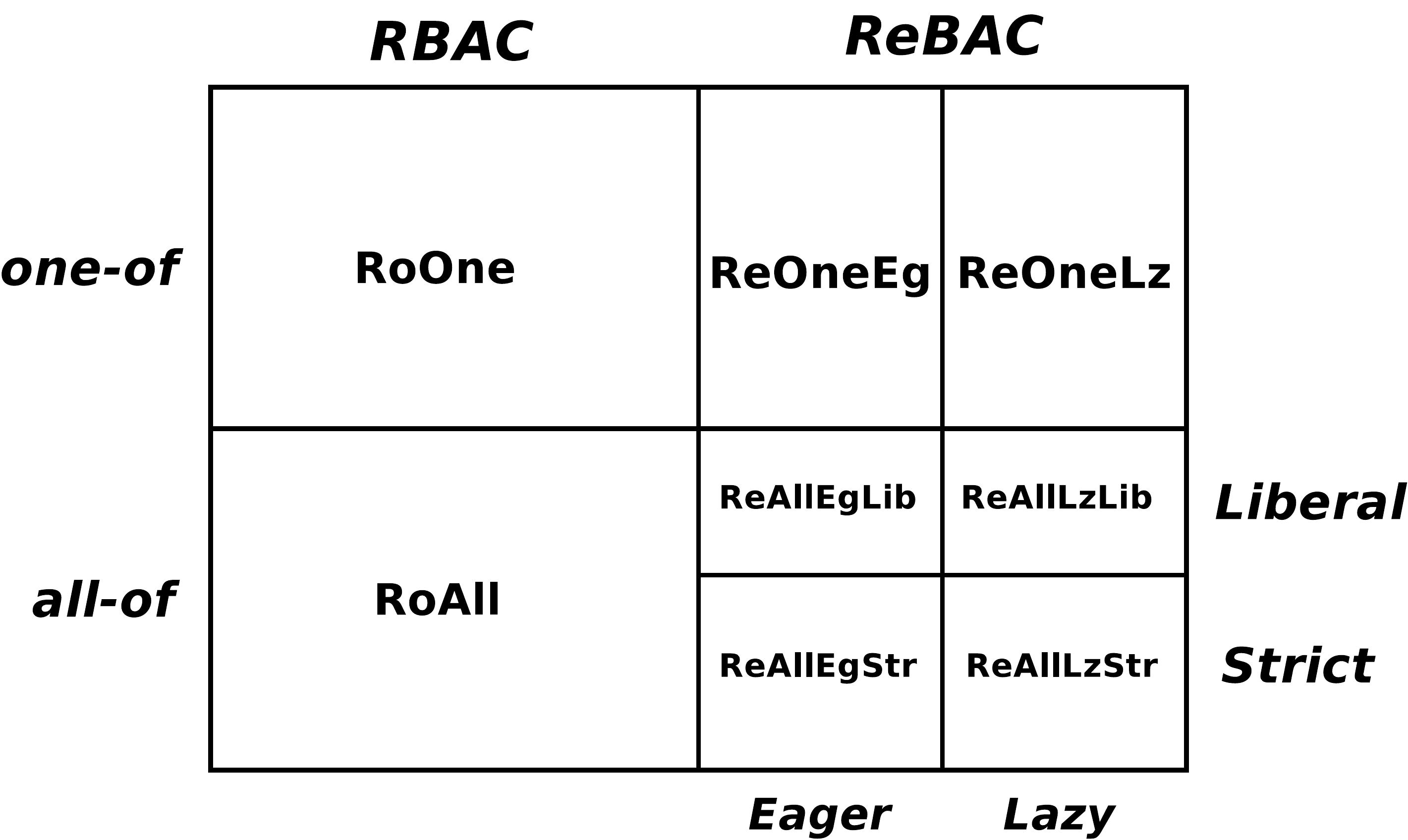}
\caption{\label{fig-configs}The 8 experimental configurations.}
\end{figure}

An empirical study has been conducted to evaluate the performance of
the various authorization schemes proposed in this paper.  As our
ReBAC-equipped version of OpenMRS is not yet deployed in any clinical
setting, no production data set is available, and thus performance
evaluation was conducted with synthetic data.  Rather than performing
``disembodied'' simulation of the various authorization schemes, we
measured the performance of those schemes within the infrastructure of
OpenMRS, thereby capturing the overhead in a realistic implementation.

We compared the performance of the OpenMRS authorization mechanism in
eight different configurations (Fig.~\ref{fig-configs}).  The two RBAC
configurations (\textsf{Ro}*) correspond to OpenMRS with only the
legacy RBAC authorization mechanism (i.e., ReBAC is turned off).
These two configurations differ in whether requests are directed
against \textcode{one-of} guards (\textsf{RoOne}) or \textcode{all-of}
guards (\textsf{RoAll}).  ReBAC authorization is turned on (and RBAC is
turned off) for the remaining six configurations
(\textsf{Re}*).  Two of the ReBAC configurations correspond to
\textcode{one-of} requests (\textsf{ReOne}*). They differ in whether
eager- or lazy-match strategy is implemented.  The last four ReBAC
configurations correspond to \textcode{all-of} requests
(\textsf{ReAll}*).  In the case of \textcode{all-of} guards, there are
two possible semantics (liberal- or strict-grant), as well as two
possible matching strategies (eager or lazy), resulting in four
configurations: \textsf{ReAllEgLib}, \textsf{ReAllEgStr},
\textsf{ReAllLzLib}, and \textsf{ReAllLzStr}.

\begin{figure}
\centering
{\small
\begin{tabular}{|l|c|l|c|l|c|} \hline
Users: & 10,000 &
Privileges: & 200 &
Roles: & 67 \\ \hline
\multicolumn{4}{|l|}{Privilege-role assignment pairs:} &
\multicolumn{2}{c|}{469}  \\ \hline
\multicolumn{4}{|l|}{User-role assignment pairs:} &
\multicolumn{2}{c|}{50,000} \\ \hline
\end{tabular}
}
\caption{\label{fig-RBAC-params}RBAC Parameters}
\end{figure}

\textit{\textbf{RBAC Protection State.}}  We randomly synthesized an
RBAC protection state for the two RBAC 
configurations (\textsf{Ro}*). OpenMRS pre-compiles the role hierarchy into a flat
space of roles.  The RBAC protection state therefore contains a
user-role assignment and a privilege-role assignment, but not a role
hierarchy.  Fig.~\ref{fig-RBAC-params} enumerates the parameters used
for synthesizing the RBAC protection state.  Justifications for the
choice of these parameters are given in Appendix \ref{app-RBAC}.

\textit{\textbf{ReBAC Protection State.}}  For the six ReBAC
configurations (\textsf{Re}*), we constructed an authorization graph
out of a social network dataset, \textsf{soc-Pokec}, obtained from the
Stanford Large Network Dataset Collection \cite{SNAP}.  The graph has
1.6 million nodes and 30 million directed edges.  This dataset is thus
even bigger than what the OpenMRS community calls a high-density
deployment.\footnote{According to a thread in the OpenMRS developer
  forum \cite{Deployment}, the number of patient records in various
  reported OpenMRS deployments ranges from 8,982 to 741,606.}

To construct the authorization graph, we identified 10,000 nodes with
the highest in-degrees, and labelled them as users (i.e.,
clinicians).\footnote{Our intuition is that clinicians are more
  connected than patients.  Specifically, they have more
  incoming edges, for example, to indicate who is the attending
  clinician of whom.}  The remaining nodes are patients.
Consequently, a directed edge in the social graph can be one of four
types: user-user, user-patient, patient-user, or patient-patient.
According to the type of each directed edge, we then randomly labelled
the directed edges using the relation identifiers of the Electronic
Health Records System case study in \cite[\S 5]{Fong:2011}.  A
detailed list of relation identifiers and the distribution of the edge
labels can be found in Appendix \ref{app-edge-lab}.

\textit{\textbf{ReBAC Policies.}}  The six ReBAC configurations
(\textsf{Re}*) presumes the existence of ReBAC policies (authorization
principals).  We generated an authorization principal for each of the
67 roles.  Authorization rules were formulated in such a way that each
principal grants the same privileges as its corresponding role.
Principal matching rules were in place so that every authorization
principal is associated with a randomly generated hybrid logic
formula.  Specifically, from the two example formulas in the
Electronic Health Records System case study in \cite[\S 5]{Fong:2011},
we extracted ten hybrid logic formulas for our experiment.  For each
principal, a formula was randomly selected from those ten formulas
(with equal probability).  See Appendix \ref{app-formulas} for
details.

\textit{\textbf{Methods, Guards, and Requests.}}  As the existing
service-layer methods of OpenMRS will not work with the authorization
graph synthesized above, we randomly synthesized service-layer methods
for the purpose of this experiment.  Each synthesized service-layer
method takes a patient as an argument, and is invoked by a user (i.e.,
clinician).  A guard is randomly generated for each method;
\textcode{one-of} guards for the *\textsf{One}* configurations, and
\textcode{all-of} guards for the *\textsf{All}* configurations.  Each
method has an empty body as we are only concerned about authorization
overhead.  For each of the eight configurations, we generated 200
method calls, with randomly selected clinicians and patients.  The
authorization times of the 200 method calls are then averaged and
reported.  Details can be found in Appendix \ref{app-requests}.


\begin{figure}
\centering
\includegraphics[width=.6\linewidth]{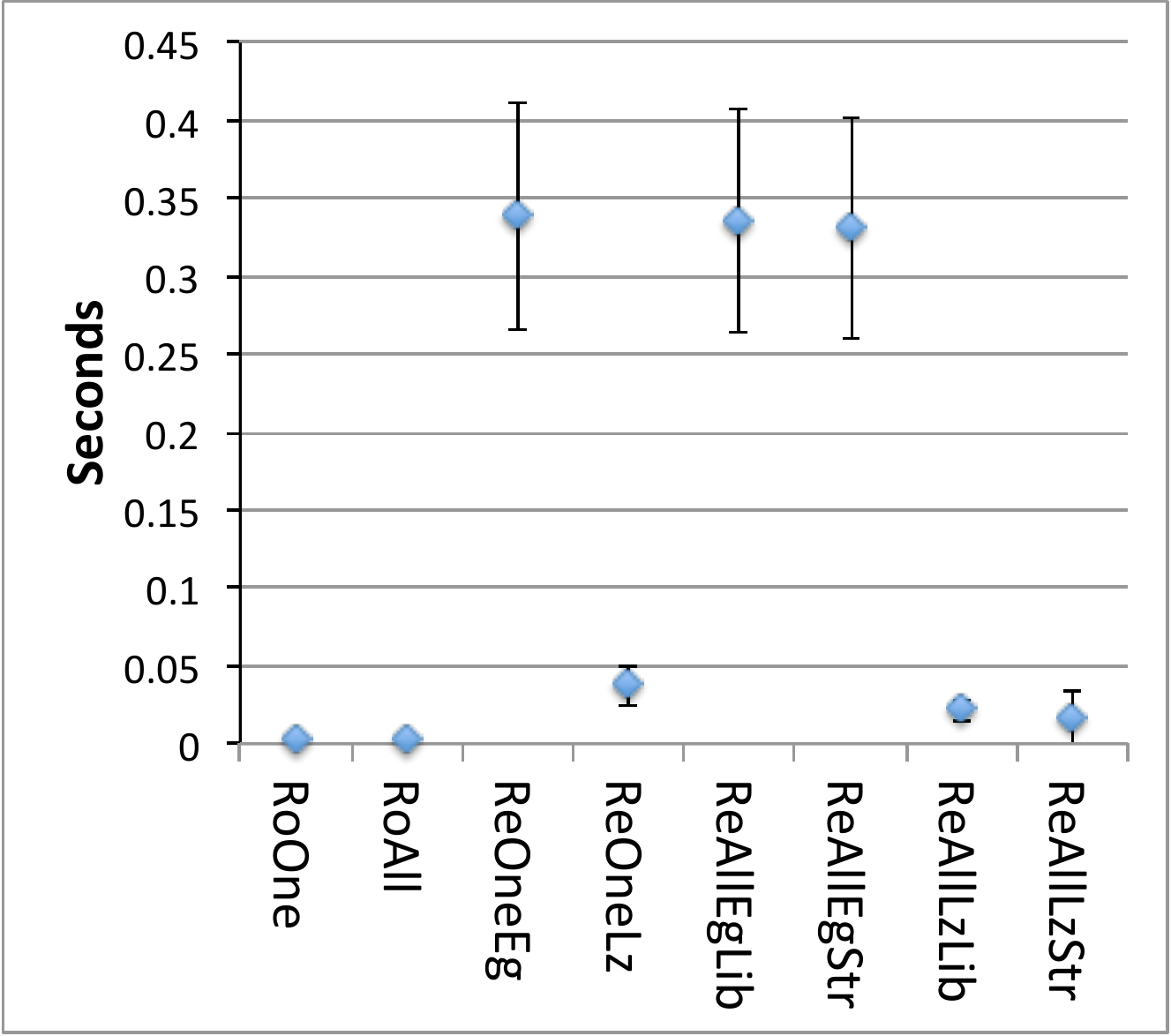}
\caption{\label{fig-time}Average time for an authorization check (with
  95\% confidence interval).}
\end{figure}

\textit{\textbf{Results and Discussions.}}  We conducted the
experiment on a desktop machine
with AMD
FX-8350 8-core Processor (16 MB cache), 16 GB RAM (1866 MHz, DDR3),
and an 840 EVO Solid State Drive running Windows 8 OS. The results are shown in
Fig.~\ref{fig-time}.


The baseline RBAC configurations (\textsf{Ro}*) incur negligible
running time.  The three eager-match configurations (*\textsf{Eg}*)
have authorization time averaging around 0.33 seconds, suggesting that
the eager-match strategy is not practical.  In contrast, all the
lazy-match configurations (*\textsf{Lz}*) have competitive
authorization times averaging around 0.016--0.037 seconds.  In
summary, matching strategy, and not authorization semantics,
is the key determinant of performance.

In our experience, the main performance overhead comes not from
backtracking within the hybrid logic model checker, but from database
accesses.  Due to the sheer size of the authorization graph (1.6
million nodes, 30 million edges), simply retrieving the neighbours of
a given node takes 0.002 second in our preliminary experiments,
resulting in unacceptable authorization times.  Noticing this, we
stored the access control relationships in a graph database (Neo4j
\cite{Neo4J}) instead of the original relational database (MySQL
\cite{MySQL}), resulting in a 20-fold speed-up in neighbor
retrieval time (0.0001 sec) and thus the fast authorization
times reported above (Fig.~\ref{fig-time}).
\begin{lesson}
  A graph database offers more competitive ReBAC authorization 
  performance than a relational database.
\end{lesson}

\section{Conclusions and Future Work}

This ReBAC adaptation of OpenMRS is the first implementation of ReBAC
in a production-scale electronic medical records system.  We reported
reusable engineering lessons for ReBAC deployment,
presented extensions of advanced
ReBAC features recently proposed by Crampton and Sellwood
\cite{Crampton-Sellwood:2014}, designed and implemented the first
administrative model for ReBAC, and evaluated the performance of
authorization checks.

Our implementation can serve as a testbed for future extensions of
ReBAC.  A number of research opportunities are motivated by this
implementation exercise.  First, the way ReBAC interacts with the
legacy RBAC mechanism is by way of conjunction: access is granted if
both access control subsystem grant access.  What are other ways in
which RBAC and ReBAC can interact with one another to deliver advanced
access control features?  Second, authorization in OpenMRS is
performed through the satisfaction of privilege requirements known as
guards.  These privilege requirements interact with the design of
other access control features (e.g., authorization principals,
authorization algorithms, positive and negative permissions, conflict
resolution) in an intimate manner.  We have opted for simplicity in
most of our design choices.  Further studies on how advanced access
control features can be implemented in the presence of OpenMRS-style
privilege requirements is a research challenge.  Third, while we have
fashioned the first administrative model for ReBAC, the theory of
ReBAC administrative models is an unexplored area.  How does one
perform, say, safety analysis in this administrative model \cite{HRU}?
Fourth, in the original proposal of ReBAC \cite{Fong:2011}
relationships are contextual.  For example, a referral relationship is
effective only in the context of a certain medical case.  Context
creation and removal provide a clean mechanism for expiration of
tentative relationships.  How does one implement contexts in OpenMRS,
especially with usability in mind?

\bibliographystyle{acm}
\bibliography{library}

\begin{thebibliography}{10}

\bibitem{Deployment}
Max number of records in an {OpenMRS} implementation.
\newblock \texttt{http://listarchives.openmrs.org/
  Max-number-of-records-in-an-OpenMRS- implementation-td4662224.html}.

\bibitem{MySQL}
{MySQL}.
\newblock \url{http://www.mysql.com/}.

\bibitem{Neo4J}
{Neo4J}.
\newblock \url{http://neo4j.com/}.

\bibitem{OpenMRS}
{OpenMRS}.
\newblock \url{http://openmrs.org/}.

\bibitem{SNAP}
Stanford {L}arge {N}etwork {D}ataset {C}ollection.
\newblock \url{http://snap.stanford.edu/data}.

\bibitem{Aktoudianakis-et-al:2013}
{\sc Aktoudianakis, E., Crampton, J., Schneider, S., Treharne, H., and Waller,
  A.}
\newblock Policy templates for relationship-based access control.
\newblock In {\em Proceedings of the 11th Annual International Conference on
  Privacy, Security and Trust (PST'13)\/} (Tarragona, Catalonia, Spain, July
  2013), IEEE, pp.~221--228.

\bibitem{Anderson}
{\sc Anderson, R.}
\newblock {\em Security Engineering}, 2nd~ed.
\newblock Wiley, 2008.

\bibitem{ReBACLib}
{\sc Anonymized}.
\newblock {ReBAC} {Java} {Library}.

\bibitem{Beznosov}
{\sc Beznosov, K.}
\newblock Requirements for access control: {US} healthcare domain.
\newblock In {\em Proceedings of the Fourth ACM Workshop on Role-Based Access
  Control (RBAC'1998)\/} (Fairfax, VA, Oct. 1998), p.~43.

\bibitem{Bruns-etal:2012}
{\sc Bruns, G., Fong, P. W.~L., Siahaan, I., and Huth, M.}
\newblock Relationship-based access control: Its expression and enforcement
  through hybrid logic.
\newblock In {\em Proceedings of the 2nd ACM Conference on Data and Application
  Security (CODASPY'12)\/} (San Antonio, TX, USA, Feb. 2012).

\bibitem{Carminati-Ferrari:2009}
{\sc Carminati, B., and Ferrari, E.}
\newblock Enforcing relationships privacy through collaborative access control
  in web-based social networks.
\newblock In {\em Proceedings of the 5th International Conference on
  Collaborative Computing: Networking, Applications and Worksharing
  (CollaborateCom'09)\/} (Washington DC, USA, Nov. 2009).

\bibitem{Carminati-etal:2009}
{\sc Carminati, B., Ferrari, E., and Perego, A.}
\newblock Enforcing access control in web-based social networks.
\newblock {\em ACM Transactions on Information and System Security 13}, 1 (Oct.
  2009).

\bibitem{Cheng-et-al:2012b}
{\sc Cheng, Y., Park, J., and Sandhu, R.}
\newblock Relationship-based access control for online social networks: Beyond
  user-to-user relationships.
\newblock In {\em Proceedings of the 4th IEEE International Conference on
  Information Privacy, Security, Risk and Trust (PASSAT'12)\/} (Amsterdam,
  Netherlands, Sept. 2012).

\bibitem{Cheng-et-al:2012a}
{\sc Cheng, Y., Park, J., and Sandhu, R.}
\newblock A user-to-user relationship-based access control model for online
  social networks.
\newblock In {\em Proceedings of the 26th Annual IFIP WG 11.3 Working
  Conference on Data and Applications Security and Privacy (DBSec'12)\/}
  (Paris, France, July 2012), vol.~7371 of {\em LNCS}.

\bibitem{Crampton:2008}
{\sc Crampton, J.}
\newblock Why we should take a second look at access control in {Unix}.
\newblock In {\em Proceedings of the 13th Nordic Workshop on Secure {IT}
  Systems (NordSec'08)\/} (Copenhagen, Denmark, Oct. 2008).

\bibitem{Crampton-Sellwood:2014}
{\sc Crampton, J., and Sellwood, J.}
\newblock Path conditions and principal matching: A new approach to access
  control.
\newblock In {\em Proceedings of the 19th ACM Symposium on Access Control
  Models and Technologies (SACMAT'14)\/} (London, Ontario, Canada, June 2014).

\bibitem{Ene-et-al:2008}
{\sc Ene, A., Horne, W.~G., Milosavljevic, N., Rao, P., Schreiber, R., and
  Tarjan, R.~E.}
\newblock Fast exact and heuristic methods for role minimization problems.
\newblock In {\em Proceedings of the 13th ACM Symposium on Access Control
  Models and Technologies (SACMAT'08)\/} (Estes Park, CO, June 2008),
  pp.~1--10.

\bibitem{Fischer-et-al:2009}
{\sc Fischer, J., Marino, D., Majumdar, R., and Millstein, T.~D.}
\newblock Fine-grained access control with object-sensitive roles.
\newblock In {\em Proceedings of the 23rd European Conference on
  Object-Oriented Programming (ECOOP'09)\/} (Genoa, Italy, July 2009),
  vol.~5653 of {\em LNCS}, pp.~173--194.

\bibitem{Fong:2011}
{\sc Fong, P. W.~L.}
\newblock Relationship-based access control: Protection model and policy
  language.
\newblock In {\em Proceedings of the First ACM Conference on Data and
  Application Security and Privacy (CODASPY'11)\/} (San Antonio, Texas, USA,
  Feb. 2011), pp.~191--202.

\bibitem{Fong-etal:2013}
{\sc Fong, P. W.~L., Mehregan, P., and Krishnan, R.}
\newblock Relational abstraction in community-based secure collaboration.
\newblock In {\em Proceedings of the 20th ACM Conference on Computer and
  Communications Security (CCS'13)\/} (Berlin, Germany, Nov. 2013),
  pp.~585--598.

\bibitem{Fong-Siahaan:2011}
{\sc Fong, P. W.~L., and Siahaan, I.}
\newblock Relationship-based access control policies and their policy
  languages.
\newblock In {\em Proceedings of the 16th ACM Symposium on Access Control
  Models and Technologies (SACMAT'11)\/} (Innsbruck, Austria, June 2011),
  pp.~51--60.

\bibitem{Gates:2007}
{\sc Gates, C.~E.}
\newblock Access control requirements for {Web} 2.0 security and privacy.
\newblock In {\em IEEE Web 2.0 privacy and security workship (W2SP'07)\/}
  (Oakland, California, USA, May 2007).

\bibitem{JavaLangSpec}
{\sc Gosling, J., Joy, B., Steele, G., Bracha, G., and Buckley, A.}
\newblock {\em The Java Language Specification}, {Java SE 8}~ed.
\newblock Oracle America, Inc, Mar. 2014.

\bibitem{HRU}
{\sc Harrison, M.~A., Ruzzo, W.~L., and Ullman, J.~D.}
\newblock Protection in operating systems.
\newblock {\em Communications of the ACM 19}, 8 (Aug. 1976), 461--471.

\bibitem{Hibernate}
Hibernate.
\newblock \url{http://hibernate.org/}.

\bibitem{Kiczales-etal:1997}
{\sc Kiczales, G., Lamping, J., Mendhekar, A., Maeda, C., Lopes, C., Loingtier,
  J.-M., and Irwin, J.}
\newblock Aspect-oriented programming.
\newblock In {\em Proceedings of the European Conference on Object-Oriented
  Programming (ECOOP'97)\/} (Jyv\"{a}skyl\"{a}, Finland, June 1997), M.~Aksit
  and S.~Matsuoka, Eds., vol.~1241 of {\em LNCS}, Springer, pp.~220--242.

\bibitem{CritiqueANSI}
{\sc Li, N., Byun, J.-W., and Bertino, E.}
\newblock A critique of the {ANSI} standard on role-based access control.
\newblock {\em IEEE Security and Privacy 5}, 6 (Nov. 2007), 41--49.

\bibitem{OpenMRSPosting}
{\sc Mamlin, B.}
\newblock Re: Policy based access control for {OpenMRS}.
\newblock
  \url{http://listarchives.openmrs.org/Policy-based-Access-Control-for-OpenMRS-td7220685.html},
  Jan. 2012.

\bibitem{Rostad-Edsberg}
{\sc R{\o}stad, L., and Edsberg, O.}
\newblock A study of access control requirements for healthcare systems based
  on audit trails from access logs.
\newblock In {\em Proceedings of the 22nd Annual Computer Security Applications
  Conference (ACSAC'2006)\/} (Miami Beach, Florida, USA, Dec. 2006),
  pp.~175--186.

\bibitem{sandhu1999arbac97}
{\sc Sandhu, R., Bhamidipati, V., and Munawer, Q.}
\newblock The {ARBAC97} model for role-based administration of roles.
\newblock {\em ACM Transactions on Information and System Security 2}, 1
  (1999), 105--135.

\bibitem{Sandhu-etal:1996}
{\sc Sandhu, R.~S., Coyne, E.~J., Feinstein, H.~L., and Youman, C.~E.}
\newblock Role-based access control models.
\newblock {\em IEEE Computer 19}, 2 (Feb. 1996), 38--47.

\bibitem{Spring}
Spring framework.
\newblock \url{http://projects.spring.io/spring-framework/}.

\bibitem{Location}
{\sc Tarameshloo, E., and Fong, P. W.~L.}
\newblock Access control models for geo-social computing systems.
\newblock In {\em Proceedings of the 19th ACM Symposium on Access Control
  Models and Technologies (SACMAT'14)\/} (London, Ontario, Canada, June 2014).

\end{thebibliography}


\appendix

\section{RBAC Protection State}
\label{app-RBAC}

We create 10,000 users for our experiments.  Since OpenMRS has 184
distinct built-in privileges, we round up and thus create 200
privileges.  We deduce several ratios from the ``healthcare'' database
of \cite{Ene-et-al:2008}: (a) role-to-privilege ratio is 1:3; (b)
average number of roles per user is 5; (c) average number of privilege
per role is 7.  From these ratios, we create 67 roles ($\approx
200/3$), 469 privilege-role assignment pairs ($\approx 67 \times 7$),
and 50,000 user-role assignment pairs ($\approx 10000 \times 5$).

\section{R\MakeLowercase{e}BAC Protection State}
\label{app-edge-lab}

From \cite[\S 5]{Fong:2011}, we extract the following relation
identifiers. We indicate below the type of each identifier: e.g., an
identifier of the patient-user type is identified by ``p-u''.
\begin{quote}
\begin{tabular}{|l|c||l|c|} \hline
Rel.~Id. & Type & Rel.~Id. & Type \\ \hline
\textsf{gp} & p-u & 
\textsf{register-ward} & p-u  \\
\textsf{referrer} & u-u &
\textsf{ward-nurse} & u-u  \\
\textsf{appoint-team} & u-u & 
\textsf{agent} & p-p \\
\textsf{team} & u-u & & \\ \hline
\end{tabular}
\end{quote}
Every directed edge in the social graph belongs to one of the four
types: user-user, user-patient, patient-user, patient-patient.  Based
on the type of a given directed edge, a relation identifier of that
type is random selected (with uniform distribution).  Note that there
is no relation identifier that has the type user-patient.  For those
edges, a dummy relation identifier is assigned.

\section{R\MakeLowercase{e}BAC Policies}
\label{app-formulas}

The Electronic Health Records System case study of \cite[\S
5]{Fong:2011} has two formulas that we can use in our experiments.
The first formula, specifying the patient-clinician relation, is
constructed incrementally in four stages in \cite[\S 5.1]{Fong:2011}.
We take the subformulas constructed in the various stages as candidate
formulas for our experiment.
\begin{quote}
\begin{tabular}{l} 
$\phi_1 = \langle \mathsf{gp} \rangle
\mathsf{requestor}$ \\
$\phi_2 = \langle \mathsf{gp} \rangle
   \langle -\mathsf{referrer} \rangle \mathsf{requestor}$\\
$\phi_3 = \phi_1 \lor \phi_2$ \\
$\phi_4 = \langle \mathsf{gp} \rangle
    \langle -\mathsf{referrer} \rangle
    \langle \mathsf{appoint\textsf{-}team} \rangle 
  \mathsf{requestor}$\\
$\phi_5 = \langle \mathsf{gp} \rangle
    \langle -\mathsf{referrer} \rangle
    \langle \mathsf{appoint\textsf{-}team} \rangle 
    ( 
  \mathsf{requestor} \lor \mbox{}$\\
  \multicolumn{1}{r}{
  $\langle \mathsf{member} 
   \rangle \mathsf{requestor})$} \\
$\phi_6 = \phi_3 \lor \phi_5$ \\
$\phi_7 = \langle \textsf{register-ward} \rangle 
  \mathsf{requestor}$\\
$\phi_8 = \langle \textsf{register-ward} \rangle 
  (\mathsf{requestor} \lor \mbox{}$\\
\multicolumn{1}{r}{
  $\langle \mathsf{ward\textsf{-}nurse} \rangle 
  \mathsf{requestor})$}\\
$\phi_9 = \phi_6 \lor \phi_8$ 
\end{tabular}
\end{quote}
The last candidate formula is basically a minor adaptation of the
formula expressing the agency relation in \cite[\S 5.2]{Fong:2011}.
\begin{quote}
\begin{tabular}{l}
$\phi_{10} = \langle \mathsf{gp} \rangle
   \mathsf{requestor} \lor 
    \langle -\mathsf{agent} \rangle
  \langle \mathsf{gp} \rangle
 \mathsf{requestor}$
\end{tabular}
\end{quote}

\section{Authorization Requests}
\label{app-requests}

We generated 400 methods with \textcode{all-of} guards, and another
400 with \textcode{one-of} guards.  The set $P$ of privileges for each
guard contains a minimum of one and a maximum of three privileges 
randomly selected from the 200 available privileges 
(Fig.~\ref{fig-RBAC-params}).\footnote{The service-layer methods of 
OpenMRS never have a privilege set of size larger three.} In addition
to the privileges we randomly generated a list of 400 clinician, and
a list of 400 participants, to serve as participants in the
authorization requests.

The methods were invoked in order (from 1 to 400) along with the
corresponding clinician, patient pair. This process was uniformly
conducted for all configurations, with the *\textsf{One}*
configurations invoking the methods with the \textcode{one-of} guards,
and the *\textsf{All}* configurations invoking the methods with the
\textcode{all-of} guards. The first 200 method invocations were
discarded as they were used for warming up the Java Virtual Machine.
The performance of the remaining 200 method invocations were recorded.

\section{N\MakeLowercase{eo}4\MakeLowercase{j} Warmup}
Retrieving the neighbourhood in Neo4J normally start out slow then 
speeds up, and stabilize at an average of 0.0001 seconds after
approximately 250 queries. Therefore, we randomly generated 250 distinct
neighbourhood retrieval queries that were ran before the method 
invocations for each test configuration.

\end{document}